# Mobile IP and protocol authentication extension

Phuc V. Nguyen, Ph.D. candidate
#127, Truong Dinh St., Ward 7, Dist. 3, Ho Chi Minh City, Vietnam
HP: +84.938201102

x201102x@gmail.com

**ABSTRACT**

Mobile IP is an open standard, defined by the Internet Engineering Task Force (IETF) RFC 3220. By using Mobile IP, you can keep the same IP address, stay connected, and maintain ongoing applications while roaming between IP networks. Mobile IP is scalable for the Internet because it is based on IP—any media that can support IP can support Mobile IP.

**I. INTRODUCTION**

In IP networks, routing is based on stationary IP addresses, similar to how a postal letter is delivered to a fixed address on an envelope. A device on a network is reachable through normal IP routing by the IP address it is assigned on the network.

However, problems occur when a device roams away from its home network and is no longer reachable using normal IP routing. This causes the active sessions of the device to be terminated. Mobile IP enables users to keep the same IP address while traveling to a different network (which may even be operated by a different wireless operator), thus ensuring that a roaming individual can continue communication without sessions or connections being dropped.

Because the mobility functions of Mobile IP are performed at the network layer rather than the physical layer, the mobile device can span different types of wireless and wire line networks while maintaining connections and ongoing applications. Remote login, remote printing, and file transfers are examples of applications where it is undesirable to interrupt communications while an individual roams across network boundaries. Furthermore, certain network services, such as software licenses and access privileges, are based on IP addresses. Changing these IP addresses can compromise the network services.

**II. THE NEED FOR MOBILE IP**

It has been foreseen that mobile computing devices will become more pervasive, more useful, and more powerful in the future. The power and usefulness will come from being able to extend and integrate the functionality of all types of communication such as Web browsing, e-mail, phone calls, information retrieval, and perhaps even video transmission. For Mobile IP computing to become as pervasive as stationary IP networks of the world, a ubiquitous protocol for the integration of voice, video, and data must be developed. The most widely researched and developed protocol is Mobile IP.

The advantages of using Mobile IP can be summarized as follows:

+ It follows fast, continuous low-cost access to corporate networks in remote areas where there is no public telephone system or cellular coverage.

+ It supports a widely range of applications from Internet access and e-mail to e-commerce.

+ Users can be permanently connected to their Internet provider and charged only for the data packets that are sent and received.

+ Lower equipment and utilization costs for those requiring reliable high-speed data connections in remote locations worldwide.

+ A user can take a palmtop or laptop computer anywhere without losing the connection to the home network.

+ Mobile IP finds local IP routers and connections automatically. It is phone-jack and wire-free.

+ Other than mobile nodes/routers, the remaining routers and hosts will still use current IP. Mobile IP leaves transport and higher protocols unaffected.

+ Authentication is performed to ensure that rights are being protected.

+ Mobile IP can move from one type of medium to another without losing connectivity. It is unique in its ability to accommodate heterogeneous mobility in addition to homogeneous mobility.

The Mobile IP network is also characterized by some disadvantages as well:

+ There is a routing inefficiency problem caused by the "triangle routing" formed by the home agent, correspond to host, and the foreign agent. It is hoped Mobile IPv6 can solve the problem.

+ Security risks are the most important problem facing Mobile IP. Besides the tradition security risks with IP, one has to worry about faked care-of addresses.

By obtaining the mobile host's care-of address and rerouting the data to itself, an attacker can obtain unauthorized information. However, another issue related to the security is how to make Mobile IP coexist with the security features coming in use within the Internet.

The characteristics that should be considered as baseline requirements to be satisfied by any candidate for Mobile IP are the following:

+ Compatibility: A new standard cannot require changes for applications or network protocols already in use. Mobile IP has to remain compatible to all lower layers used for the standard non-mobile IP. It must not require special media or protocol.

+ Transparency: Mobility should remain "invisible" for many higher layers protocols and applications. Besides maybe noticing a lower bandwidth and some interruption in service, higher layers should continue to work, even if the mobile changed its point of attachment to the network.

+ Scalability and efficiency: Introducing a new mechanism into the Internet must not degrade the efficiency of the network. Due to the growth rates of mobile communication, clearly Mobile IP must be scalable over a large number of participants in the whole Internet.

+ Security: All messages used to transmit information to another node about the location of a mobile node must be authentication to protect against remote redirection attacks.

The requirements of Mobile IP may be summarized as follows:

+ A mobile node must be able to communicate with other nodes after changing its link-layer point of attachment to the Internet, yet without changing its IP address.

+ Application programs must be able to operate continuously over a single session while the network attachment point of the mobile host changes.

+ A mobile node must be able to communicate with other nodes that do not implement these mobility functions.

+ All messages used to update another node with the location of a mobile node must be authenticated in order to protect against redirection attacks.

### III. PROTOCOL AUTHENTICATION EXTENSION

In a routing protocol, it is important for peers to trust one another and to ensure that the messages they exchange have not been altered in transit. False routing updates from un-trusted peers or altered updates from trusted peers can wreak havoc on a network, for example, causing traffic for multiple prefixes to be routed incorrectly or even black holed.

The Mobile IP protocol uses security authentication extensions to provide peer authentication and message integrity. As the name suggests, the authentication extension is an extension containing relevant security information that is appended to the end of the message. The authentication extensions are designed to allow extensive flexibility through their extension type and placement within the message. That is, different types of authentication extension's secure messages between different Mobile IP entities.

Moreover, the authentication extensions secure a specific part of the Mobile IP messages, depending on where the part is placed within the message.

The critical purpose of the authentication extension is to verify the sender of the message and to ensure that the message was not altered while in transit. The extension types are allocated to allow authentication between various pairs of peers. The following four Mobile IP extension types, allocated for authentication, currently exist:

+ Mobile Node - Home Agent Authentication Extension – MHAE.

+ Mobile Node - FA Authentication Extension – MFAE.

+ FA - Home Agent Authentication Extension – FHAE.

+ Generalized Authentication Extension – GNAE.

Authentication extensions allow flexibility in authenticating various parts of the registration message with various peers. Thus, as information is added to the registration control messages, it can be protected without altering the protection to pre-existing portions of the control message. For example, a Mobile Node can secure its RRQ with an MHAE and forward the request to its FA. The MHAE protects the information preceding it, namely, the base RRQ. In turn, the FA might want to append an extension to the RRQ and secure the extension with an FHAE. The two extensions secure different subsets of the message and are between different Mobile IP entity pairs. That is, the base RRQ is secured between the Mobile Node and Home Agent, and the entire message, including the appended extension, is secured between the FA and Home Agent.

### 3.1. Security Associations

For Mobile IP entities to use an authentication extension between them, they must first share a security relationship. This relationship is in the form of a set of predefined parameters configured into each node, which are collectively known as a security context. A security context is comprised of the following components:

+ Algorithm and mode to be used in crypto computations.

+ Shared key between the peers.

+ Replay protection method.

Theoretically, each node can have more than 4 billion individual security contexts per peer. A group of one or more security contexts that are shared with an individual peer is referred to as a security association. It is also common for the term security association to be used interchangeably with the term security context. This is likely because security associations are often made up of only one security context.

The sender computes a cryptographic keyed hash of the message using an algorithm and shared key, places this value in the authentication field of the authentication extension, and sends the message. The algorithm and keys that protect a message are implied by the sender within the authentication extension, which is part of the protected portion of the message. To verify the integrity of a message, the recipient computes its own cryptographic keyed hash of the same portion of the message (the message not including the authenticator value) using the same algorithm and shared key.

It then compares the computed hash to the authenticator value in the appropriate authentication extension. If the two match, the message is considered to be authenticated.

### 3.2. SPI

A specific security context is identified in the authentication extension by the security parameter index (SPI) value. The SPI is a 4-byte value that is configured as either a hexadecimal or decimal value. (Unfortunately, this can often lead to confusion because the peer devices can require the SPI value to be specified in different formats.) If a security violation is received on a RRQ, the SPI value is the first item that should be verified, because it identifies the security context to use in authenticating the message.

### 3.3. Algorithm and Mode

The authenticator value in the authentication extension is a message authentication code (MAC), which can be thought of as a fingerprint. For each registration message, a hash algorithm calculates the unique fingerprint value, which is of smaller total length than the original message. Thus, given that more possible values exist for the original message than do unique fingerprint results, the algorithms are designed so that the result is as unique as possible. The ideal algorithm results in 50 percent of the bits in the hash changing by changing just 1 bit in the message.

### 3.4. Key

The authenticator is computed using a key that is shared between the two peers. The node that initiates the message uses the key to compute the value in the authentication extension. The node that receives the message uses the same key value and computes the authenticator over the registration message and compares the result to the value in the authentication extension. If the two values match, the authentication is accepted. The default key is 128 bits long and is usually represented by 32 hex characters.

RFC 3344, "IP Mobility Support for IPv4," states the following:

The default algorithm is HMAC-MD5 [23], with a key size of 128 bits. The FA MUST also support authentication using HMAC-MD5 and key sizes of 128 bits or greater, with manual key distribution. Keys with arbitrary binary values MUST be supported.